# Entropy Gain and Information Loss by Measurements

Xing M. Wang[1]

## Abstract

When the von Neumann entropy (VNE) of a system increases due to measurements, certain information is lost, some of which may be recoverable. We define information retrievability (IR) and information loss (IL) as functions of the density matrix through VNE to illustrate the relationship between gain and loss. We demonstrate that when a pure, unbiased m-qubit state collapses into a maximally mixed state, it experiences the maximal loss of information and the highest gain in entropy, equivalent to the m-bit classical Shannon entropy. We analyze the VNE, IR, and IL of single qubits, entangled photon pairs in Bell tests, three-qubit systems in quantum teleportation, multiple-qubit systems of GHZ and W states, and two-qubit Werner mixed states, emphasizing their IL dependence on parameters such as polarization bias and qubit count. Data exchange between two observers in Bell tests can recover some of the lost quantum information and eliminate the associated quantum entropy, even years later. The need to recover knowledge explains why no spooky action occurs at a distance. We show that measuring the Bell, GHZ, and marginally entangled Werner states yields the same minimum entropy gain (ln2) and equal minimal information loss (50 percent).

## 1. Introduction

All gates representing unitary transformations are reversible in a quantum circuit [1, p. 251]. Assume the state of a qubit in a quantum circuit is represented by

$$| q\rangle = \sqrt{2/5}\,| 0\rangle + \exp(-i4\pi/5)\sqrt{3/5}\,| 1\rangle \tag{1.1}$$

When this state is measured in the computational basis (CB), it collapses to a definite state—either $|0\rangle$ or $|1\rangle$, with the following Probability Mass Functions (PMF, [2, p4]):

$$P(0) = 2/5, \quad p(1) = 3/5 \tag{1.2}$$

In Probability Bracket Notation (PBN, [3]), the PMF can be expressed as a system P-ket:

$$| \Omega_q) = (2/5)\,| 0) + (3/5)\,| 1) \tag{1.3}$$

The measurement process results in an irreversible loss of some original information. We cannot recover Equation (1.1) from the resulting MPF because the relative phase factor [4, p.93], exp $(-i4\pi/5)$, is completely lost. How much information about the system is lost during the measurement? How do we quantify this loss? How much entropy is gained?

After discussing one-cubit cases in quantum computing, we propose Information Retrievability (IR) and Information Loss (IL) as functions of the density matrix ([1, p53]) through the change of von Neumann Entropy (VNE) [1, 179]. When a pure, unbiased 10-qubit state collapses into a maximally mixed state, it has the maximal loss of information (99.9%) and the highest gain in entropy, equivalent to the 10-bit classical Shannon entropy. Suppose a system's entropy jumps to millions of bits by certain measurements; then, the information loss approaches 100 percent, which does not equate to a loss of millions of information bits (see Section 3.4).

[1] Sherman Visual Lab, Sunnyvale, CA 94085, USA; xmwang@shermanlab.com

Next, general one-qubit pure states are investigated. Its entropy gain depends on the polarization bias related to its polar angle in the Bloch sphere [1, p.54]. We investigate the information loss of Bell states [4, p.25] in Bell tests [4, p111] [5] with mutual bias depending on the angle between the two remote devices. We demonstrate that some lost information can be recovered by exchanging knowledge between the two observers, similar to delayed choice experiments [6,7], which means a portion of the gained quantum entropy is removable.

Later, several systems of three [9] and more entangled qubits are addressed, such as the 3-qubit teleportation [4, p.26] [9], the *m*-qubit GHZ [10], and W states [10]. We address the entropy uncertainty issue in the Bell test and teleportation [11,12]. We explain why the W states are more robust than the GHZ states by comparing their entanglement information loss caused by measuring. We then explore the 2-qubit mixed Werner State [10].

Our results show that after measuring one pair of entangled qubits, the Bell, GHZ, and certain marginally entangled Werner states have an equal minimal entanglement entropy gain (ln2, irremovable) and the same minimal entanglement info loss (50%, unrecoverable).

We end our article with a summary and discussion.

## 2. Information Loss by Measurements of Single Qubit in Quantum Computing

Before going to a system of multiple qubits, let us start with measurements of a single qubit. Here and from now on, we will alternatively use the Computational Basis (CB) $|0\rangle$ and $|1\rangle$, the spin basis $|\uparrow_z\rangle$ and $|\downarrow_z\rangle$ for spin ½ particles, and the polarization basis $|H\rangle$ and $|V\rangle$ of photons (Quantum Tomography [13]) as our standard basis:

$$|0\rangle = \begin{bmatrix} 1 \\ 0 \end{bmatrix} \text{ (same as } |\uparrow_z\rangle \text{ or } |H\rangle), \qquad |1\rangle = \begin{bmatrix} 0 \\ 1 \end{bmatrix} \text{(same as } |\downarrow_z\rangle \text{ or } |V\rangle) \tag{2.1}$$

The two CB states form a complete set of orthonormal vectors in the Hilbert space $\mathcal{H}^2$[1, p37], which can readily be mapped to the P-basis in PBN [4, Eq. (19)]:

$$\hat{I}_{CB} = \sum_{\alpha=1}^{2} |\chi_\alpha\rangle\langle\chi_\alpha| = |0\rangle\langle 0| + |1\rangle\langle 1|, \quad \langle\chi_\alpha|\chi_\beta\rangle = \delta_{\alpha\beta} \tag{2.2}$$

$$\hat{I}_{CB} = \sum_{\alpha=1}^{2} |\chi_\alpha) P(\chi_\alpha| = |0)P(0| + |1)P(1|, \quad P(\chi_\alpha|\chi_\beta) = \delta_{\alpha\beta} \tag{2.3}$$

First, let us assume that the initial state of the qubit is in a pure CB state:

$$|\Psi_z\rangle = |0\rangle = |\uparrow_z\rangle \qquad \text{(an eigenstate of } \hat{S}_z \text{ or Pauli matrix } \sigma_z \text{ [1, p. 44])} \tag{2.4}$$

Using Equation (2.2-3) and the PBN, we can find the probability mass function (PMF) from Eq. (2.4) as:

$$1 = \langle \Psi_z | \Psi_z \rangle = \langle \Psi_z | \hat{I}_{CB} | \Psi_z \rangle = \langle \Psi_z | 0 \rangle \langle 0 | \Psi_z \rangle + | \langle \Psi_z | 1 \rangle \langle 1 | \Psi_z \rangle \triangleq \sum_{\alpha=1}^{2} P(\chi_\alpha | \Omega_{z,CB}) \quad (2.5)$$

$$P(\chi_\alpha | \Omega_{z,CB}) = | \langle \chi_\alpha | \Psi_z \rangle |^2, \quad P(0 | \Omega_{z,CB}) = 1, \quad P(1 | \Omega_{z,CB}) = 0, \quad \chi_\alpha \in \Omega_{z,CB} = \{0,1\} \quad (2.6)$$

This means that if we measure state Equation (2.4) in CB, we will always (100%) get the same output 0, corresponding to state $|0\rangle$. We can write its PMF-induced quantum state as:

$$| \Psi_\varphi \rangle = e^{i\varphi} | 0 \rangle \quad (2.7)$$

Because normalization can ignore the phase factor in (2.7), the induced quantum state is identical to the original state, and no information is lost; there is no state collapse. Next, we assume that the qubit has the following initial state:

$$| \Psi_{-x} \rangle = (1/\sqrt{2}) (|0\rangle - |1\rangle) \triangleq |\downarrow_x\rangle \quad \text{(an eigenstate of } \hat{S}_x \text{ or Pauli matrix } \sigma_x) \quad (2.8)$$

As expressed in the PBN, its PMF is evenly distributed (no bias):

$$P(\chi_\alpha | \Omega_{x,CB}) = | \langle \chi_\alpha | \Psi_{-x} \rangle |^2, \quad P(0 | \Omega_{x,CB}) = P(1 | \Omega_{x,CB}) = 1/2, \quad \chi_\alpha \in \Omega_{x,CB} = \{0,1\} \quad (2.9)$$

After measurements, however, the phase shift factor ($e^{i\pi} = -1$) in Equation (8), vital to quantum interference [14], is completely lost. The following PMF-induced state will produce the same PMF as in Equation (9):

$$| \Psi_\varphi \rangle = (1/\sqrt{2}) (|0\rangle + e^{i\varphi}|1\rangle) \quad (2.10)$$

According to the Copenhagen Interpretation [15], a measurement causes state collapse (unless the initial state is a measurement basis vector). It is irreversible and non-unitary, and "when there is an irreversible process, e.g., an irreversible operation or a decoherence process, in the erasure process, the information would be erased perpetually" [16].

Loss of info means more chaos or entropy, leading us to our next section.

## 3. Information Loss, Density Matrix, and von Neumann Entropy

How can we quantitatively represent the information loss due to measurements? We already know that information loss is accompanied by increased chaos or entropy.

### 3.1. The density matrix and von Neumann entropy:

For a quantum system, the von Neumann entropy (VNE [1, p.179]) is given by:

$$S(\rho) = -\text{Tr}(\rho \ln \rho) \quad (3.1)$$

Here $\rho$ is the density matrix [1, p.53], defined in Hilbert space $\mathcal{H}^N$, having the following properties:

$$\begin{aligned} &\text{Tr}\,\rho = 1 \text{ (unit trace)}, \\ &\rho = \rho^\dagger \text{ (Hermitian)}, \\ &\langle \varphi | \rho | \varphi \rangle \geq 0 \;\forall\; |\varphi\rangle \in \mathcal{H}^N \text{ (positive definite)} \end{aligned} \quad (3.2)$$

A density matrix $\rho$ describes a pure state [1, p.54] if and only if $\rho^2 = \rho$, i.e., the state is idempotent; it describes a mixed state [1, p.63] if $\rho^2 \neq \rho$ [17]. $\mathrm{Tr}\rho^2$ indicates the purity of a state. For a pure state, $\mathrm{Tr}\rho^2 = 1$, and for a mixed state, $\mathrm{Tr}\rho^2 < 1$. It can be shown that:

$S(\rho) \geq 0$; $S(\rho) = 0$ if only if $\rho$ describes a pure state.

Let $\eta_k$, $k \in \{1,...,N\}$, be the eigenvalues of $\rho$ for orthonormal eigenstate $|\eta_k\rangle$, then:

$$\rho\,|\eta_k\rangle = \eta_k\,|\eta_k\rangle, \quad \rho = \sum_k^N |\eta_k\rangle\,\eta_k\langle\eta_k\,|, \quad \sum_k^N \eta_k = 1 \tag{3.3}$$

In the orthonormal basis, the VNE is given by:

$$S(\rho) = -\mathrm{Tr}\,(\rho \ln \rho) = -\sum_{k=1} \eta_k \ln \eta_k = \ln\left(\prod_{k=1} \eta_k^{\;-\eta_k}\right) \tag{3.4}$$

The maximally mixed state (MMS [1, p.65]) in $N$-dimensional Hilbert space $\mathcal{H}^N$ has a density matrix proportional to the $N$-d identity with the maximal entropy [1, p.65]:

$$\rho_{\max} = \frac{1}{N}\hat{I}_N\,, \quad S(\rho_{\max}) = \ln N \tag{3.5}$$

Our initial state $|\Psi_{-x}\rangle$ in Eq. (2.8) is pure; therefore, its VNE is zero:

$$\rho_{in} = |\,\Psi_{-x}\rangle\langle\Psi_{-x}\,|\,, \quad \rho_{in}^{\;2} = \rho_{in}, \quad S(\rho_{in}) = 0 \tag{3.6}$$

The explicit expression of $\rho_{in}$ can be found by using Equation (2):

$$\rho_{in} = |\,\Psi_{-x}\rangle\langle\Psi_{-x}\,| = \frac{1}{2}\big(|\,0\rangle - |\,1\rangle\big)\big(\langle 0\,| - \langle 1\,|\big) = \frac{1}{2}\begin{pmatrix} 1 & -1 \\ -1 & 1 \end{pmatrix} \tag{3.7}$$

Its eigenvalues and corresponding normalized eigenvectors are:

$$\eta_1 = 1,\ |\,\eta_1\rangle = \frac{1}{\sqrt{2}}\begin{bmatrix} 1 \\ -1 \end{bmatrix} = |\,\Psi_{-x}\rangle, \quad \eta_2 = 0,\ |\,\eta_2\rangle = \frac{1}{\sqrt{2}}\begin{bmatrix} 1 \\ 1 \end{bmatrix} = |\,\Psi_{x}\rangle \tag{3.8}$$

Therefore $S(\rho_{in}) = -\mathrm{Tr}\,(\rho_{in} \log \rho_{in}) = -\sum_{k=1}^{2} \eta_k \ln \eta_k = 0$, as expected for a pure state (note that $\lim_{x \downarrow 0} x \ln x = 0$). Now, let us calculate the density matrix and its entropy after measurements in CB. From the PMF obtained in Eq. (10), we have:

$$\rho_{f,CB} = \sum_{\alpha=1}^{2} |\,\chi_\alpha\rangle\,\eta_\alpha\langle\chi_{,\alpha}\,| = |\,0\rangle\frac{1}{2}\langle 0\,| + |\,1\rangle\frac{1}{2}\langle 1|, \quad \therefore\ \rho_{f,CB} = \begin{pmatrix} 1/2 & 0 \\ 0 & 1/2 \end{pmatrix}$$

$$\therefore\ S(\rho_{f,CB}) = -\mathrm{Tr}\,(\rho_{f,CB} \ln \rho_{f,CB}) = -\sum_{\alpha=1}^{2} \eta_\alpha \ln \eta_\alpha = -\sum_{\alpha=1}^{2} \frac{1}{2} \ln \frac{1}{2} = \ln 2 \tag{3.9}$$

Since this is a 2-d Hilbert space, according to Eq. (3.5), state $\rho_{f,CB}$ represents a MMS. The entropy of a given density matrix $\rho$, defined by (3.3), can also be thought of as the average chaos of each eigenvalue of $\rho$, given by $-\ln\eta_k$ :

$$S(\rho) = -\mathrm{Tr}\,(\rho\ln\rho) = \sum_{k=1}\eta_k(-\ln\eta_k) \triangleq \langle -\ln\rho\rangle \triangleq \langle -\ln\eta\rangle \tag{3.10}$$

In the PBN, the density matrix Eq. (3.3) is equivalent to a system P-ket:

$$|\,\Omega) = \sum_{k}^{N}\ \eta_k\,|\,\eta_k), \quad P(\Omega\,|\,\Omega) = \sum_{k}^{N}\eta_k = 1 \tag{3.11}$$

Then, the VNE can be expressed as the expected value of ln($\rho$) as shown in Eq. (3.10):

$$S(\rho) = -P(\Omega\,|\ln\rho\,|\,\Omega) = -\sum_{k=1}P(\eta_k\,|\,\rho\ln\rho\,|\,\eta_k) = -\sum_{k=1}\eta_k\ln\eta_k = \ln\left(\prod_{k=1}\eta_k{}^{-\eta_k}\right) \tag{3.12}$$

### 3.2. The information retrievability and the information loss:

The increase of entropy means the rise of chaos, the decrease of order, or the loss of knowledge. How do we use it to describe the information loss numerically?

Recall that $S(\rho)\geq 0$, and more entropy means more information loss. On the other hand, a pure state has zero entropy, $S$ = 0, so it has the maximally available information, or the full knowledge, allowing us to retrieve it through quantum mechanics. The information retrievability (IR) of a state with entropy $S$ is denoted by $iR$(S). We should expect $iR$(0) = 1, an increasing $S$ leading to a decreasing $iR$(S), and $iR(\infty)$ = 0. Similar to any decay process, it is natural to define the relative change of $iR$($S$) as:

$$\frac{\Delta(iR)}{iR} = -\Delta S \ \Rightarrow\ \Delta(\ln iR) = -\Delta S$$
$$\therefore\ iR(S) = e^{-S}, \quad iR(0) = 1 \tag{3.12}$$

Let $iR$ ($S_i$) be the IR of the initial density matrix $\rho_i$, $iR$ ($S_f$) be the IR of the final $\rho_f$, then the ratio of the two IRs can be written as a two-argument function, called the comparative IR:

$$iR(S_f, S_i) \triangleq \frac{iR(S_f)}{iR(S_i)} = \frac{e^{-S_f}}{e^{-S_{in}}} = e^{S_{in}-S_f} = e^{-\Delta S_{f,i}} = iR(\Delta S_{f,i}) \tag{3.13}$$

The information unavailable after measurements is lost, so information loss (IL) is logically defined by:

$$iL(S_f, S_i) \triangleq 1 - iR(S_f, S_i) = 1 - iR(\Delta S_{f,i}) \tag{3.14}$$

In most cases of our study, the initial state is a pure state with zero entropy; therefore:

$$iR(S) \triangleq iR(S,0) = e^{-S(\rho)} \underset{(3.4)}{=} \prod\nolimits_{k=1}^{N}\eta_k{}^{\eta_k}\ ,\ iL(S) \triangleq 1 - iR(S) = 1 - \prod\nolimits_{k=1}^{N}\eta_k{}^{\eta_k} \tag{3.15}$$

For our example of $\rho_{f,CB}$, Eq. (3.9), we have:

$$iR(S_{f,CB},0)=iR(S_{CB})=\prod_{j=1}^{2}\left(1/2\right)^{1/2}=e^{-S_{CB}}=e^{-\ln 2}=½,\ iL(S_{CB})=1-iR(S_{CB})=½. \tag{3.16}$$

We lost half the initially available information after measurements in CB. If the base of the logarithm is 2, as commonly used for Shannon Entropy [1, p168], we have:

$$S_2(\rho)\triangleq-\mathrm{Tr}\,(\rho\log_2\rho)=\sum_{k=1}^{N}\log_2(\eta_k{}^{-\eta_k}),\quad iR(S_2)\triangleq 2^{-S_2(\rho)}=\prod_{k=1}^{N}\eta_k{}^{\eta_k} \tag{3.17}$$

It is identical to Eq. (3.16). Hence, by our definition, the retrievable (or lost) information is independent of the logarithm base in entropy (VNE or Shannon). Table 3.1 lists some characteristic values of VNE vs. IR and IL defined in Equations (3.13-14).

| $\Delta S(\rho)$ | 0 | ln 2 | 2 ln 2 | m ln 2 | 10 ln2 |
|---|---|---|---|---|---|
| IR ($iR$) | 1 | ½ = 50% | ¼ = 25% | $(1/2)^m$ | 0.001 |
| IL ($iL$) | 0 | ½ = 50% | ¾ = 75% | $1-(1/2)^m$ | 99.9% |

Table 3.1: Several typical values of $\Delta S(\rho)$ vs IR & IL in Eq. (1.4)

### 3.3. The maximal mixed states (MMS):

The Hilbert space of the $m$-qubit system has dimension $N = 2^m$. The density matrix of an $m$-qubit MMS in Eq. (3.5) can be rewritten as:

$$\rho_{2^m,\max}=2^{-m}I_{2^m}=\sum_{k=1}^{m}|\eta_k\rangle 2^{-m}\langle\eta_k|,\quad \{|\eta_k\rangle\}=\{|00...0\rangle...|01...1\rangle...\} \tag{3.18}$$

Similar to Eq. (2.10), the PMF-induced $m$-q pure state can be written as:

$$|\Psi_{\vec{\varphi}}\rangle=\sqrt{2^{-m}}\sum_{k=1}^{2^m}e^{i\varphi_k}\,|\eta_k\rangle \tag{3.19}$$

An $m$-q MMS has maximal entropy, minimal retrievability, and maximal info loss as follows:

$$S_{2^m,\max}=\ln 2^m=m\ln 2,\ \ iR(S_{2^m,\max})=\prod_{k=1}^{2^m}\left(2^{-m}\right)^{2^{-m}}=2^{-m},\quad iL(S_{2^m,\max})=1-2^{-m} \tag{3.20}$$

After measurements, all relative phase factors in Equation (3.19) are completely lost. The PMF describes a discrete uniform distribution (DUD) of $N = 2^m$ equal-chance outcomes (no bias) with the following base-2 classical Shannon Entropy [1, p.168]:

$$H_2(X)=-\sum_{i=1}^{2^m}p(x_i)\log_2 p(x_i)=-\sum_{i=1}^{2^m}2^{-m}\log_2 2^{-m}=m \tag{3.21}$$

If you toss $m$ fair coins infinitely many times, you will get the same uniform PMF, producing $m$-bit (or $m$-shannon) classical information content [1, p.199]. Therefore, when an $m$-q state collapses to an MMS through measurements, its entropy converts from von Neumann's quantum entropy to Shannon's $m$-bit classical information. All quantum information is lost. Eq. (3.20-21) also tells us that an $m$-qubit system at most has $m$-bit quantum information.

**3.4. The wave-particle duality and the Bekenstein bound of entropy**: Although this article concentrates on systems of qubits, measurements that cause entropy gain are not just for qubits. Assume a hydrogen atom is freely moving in a sealed long tube with a length of $L$. It has a stationary normalized pure state in the $x$-direction ($-L/2 \le x \le L/2$):

$$\langle\psi_n \mid \Psi\rangle = c_n; \qquad \langle x \mid \psi_n\rangle \equiv \sqrt{\frac{2}{L}}\cos(\frac{n\pi}{L}x) \equiv \sqrt{\frac{2}{L}}\cos(k_n x); \quad \langle\Psi \mid \Psi\rangle = \sum_{n=1}^{\infty} c_n^{\ 2} = 1 \tag{3.22}$$

Its density matrix is given by:

$$\rho = \mid \Psi\rangle\langle\Psi \mid = \sum_{m,n} \mid \psi_m\rangle\langle\psi_m \mid \Psi\rangle\langle\Psi \mid \psi_n\rangle\langle\psi_n \mid = \sum_{m,n} \mid \psi_m\rangle c_m\, c_n^{*}\langle\psi_n \mid \tag{3.23}$$

It is easy to check that Tr ρ = 1 and $\rho^2 = \rho$. As a pure state, it has zero quantum entropy. Suppose, after measuring its location in the middle of the tube, the wave function collapses to $x = 0$ and becomes Φ(x) = δ($x$). The final state |Φ⟩ cannot be normalized and is not a pure state:

$$\rho_f^{\ 2} = \mid \Phi\rangle\langle\Phi \mid \Phi\rangle\langle\Phi \mid = \mid \Phi\rangle\int_{-L/2}^{L/2} dx\langle\Phi \mid x\rangle\langle x \mid \Phi\rangle\langle\Phi \mid = \mid \Phi\rangle\int dx\,\delta^2(x)\langle\Phi \mid \neq \rho_f \tag{3.25}$$

However, we can expand state |Φ⟩ on the basis used in Eq. (3.22):

$$d_n = \langle\psi_n \mid \Phi\rangle = \int_{-L/2}^{L/2} dx\sqrt{\frac{2}{L}}\cos(k_n x)\delta(x) = \sqrt{\frac{2}{L}} \tag{3.26}$$

It represents a uniform distribution in the $k$-space ($k = p/\hbar$): the exact position leads to completely uncertain momentum! This is due to the wave-particle duality associated with the well-known Heisenberg uncertainty [18, p. 128], which gives the lower bound:

$$\Delta x\,\Delta k \geq 1/2 \tag{3.27}$$

Certainly, the precise position δ(x) and infinite entropy exist solely in thought experiments, and there is a Bekenstein upper bound for entropy [19-20]. For example, a hydrogen atom's upper entropy bound is approximately a million bits:

$$\frac{S}{\ln 2} \leq \frac{2\pi k_B RE}{\hbar c \ln 2} \approx \frac{6.28\times10^{-23}\times5\times10^{-11}\times1.5\times10^{-10}}{1.05\times10^{-34}\times3\times10^{8}\ln 2}\frac{J}{K} \approx \frac{2.18\times10^{-17}}{\ln 2}\frac{J}{K} \approx 3\times10^{6}\,\text{bits} \tag{3.28}$$

where $R$ is the radius of a sphere that can enclose the given system, and $E$ is the total energy, including any rest masses. Therefore, as an open system, its entropy can reach millions of bits. Does this mean the original finite system lost millions of bits of information? Did it contain so much information before the measurement? Our definition of information loss in Eqs (3.13-14) seems more appropriate for such cases.

## 4. Information Loss, Polarization Bias, and Realized Density Matrix of 1-q State

The realized density matrix determines the actual information loss or entropy gain; the initial state and the chosen measurement basis influence it. Assuming that the initial state is Eq. (2.8) and we choose to measure the value of $\sigma_x$, then we obtain the following PMF:

$$P(\chi_{x,\alpha} \mid \Omega_x) = \mid\langle\chi_{x,\alpha} \mid \Psi_{-x}\rangle\mid^2,\ P(\uparrow_x \mid \Omega_x) = 0,\ P(\downarrow_x \mid \Omega_x) = 1,\ \Omega_x = \{\uparrow_x, \downarrow_x\} \tag{4.1}$$

We will always yield the outcome $\sigma_x = -1$ by repeating execution. The resulting density matrix has zero quantum entropy and zero information loss.

$$\rho = \sum_{\alpha=1}^{2} |\,\chi_{x,\alpha}\rangle \eta_{x,\alpha} \langle \chi_{x,\alpha}\,| = |\downarrow_x\rangle\langle\downarrow_x| = \begin{pmatrix} 1 & 0 \\ 0 & 0 \end{pmatrix}$$

$$S(\rho_f) = -\mathrm{Tr}\,(\rho_f \log \rho_f) = 0 = S(\rho_i) \tag{4.2}$$

Therefore, if the initial state is a measurement basis state, there is no loss of information, and entropy will not increase.

$$iR(S_{f,x}, 0) = iR(0,0) = iR(0) = 1, \quad iL(S_{f,x}, 0) = 0 \tag{4.3}$$

Now, let us produce a general qubit state by rotating $|0\rangle$ with Euler polar angle $\theta$ and azimuthal angle $\varphi$ in the Bloch sphere, as shown in Fig. (4.1a):

$$|q\rangle = \cos\left(\frac{\theta}{2}\right)|\,0\rangle + \sin\left(\frac{\theta}{2}\right)e^{i\varphi}\,|\,1\rangle \tag{4.4}$$

We measure it in CB (the basis of $\sigma_z$), recording whether the qubit is 0 or 1. After repeating many, many times, we will approach the following realized density matrix from Eq. (36):

$$\rho_x = |\,0\rangle(1-x)\langle 0\,| + |\,1\rangle x \langle 1\,|, \quad x \triangleq \sin^2\left(\frac{\theta}{2}\right) \tag{4.5}$$

The corresponding VNE, IR, and IL are given by Eq. (37) and (26):

$$S(\rho_x) = -\mathrm{Tr}\,(\rho_x \ln \rho_x) = -(1-x)\ln(1-x) - x\ln x \tag{4.6}$$

$$iR(S(\rho_x) = x^x(1-x)^{(1-x)}, \; iL(S(\rho_x) = 1 - x^x(1-x)^{(1-x)} \tag{4.7}$$

To better describe their dependence on the polar angle $\theta$, we introduce the quantum polar bias $\beta$, an indicator of probability bias associated with the polar angle $\theta$ in the initial state:

$$\beta(\theta) \triangleq [\cos^2(\theta/2) - \sin^2(\theta/2)]^2 = \cos^2(\theta) = \cos^2(2\sin^{-1}\sqrt{x}) \in [0,1] \tag{4.8}$$

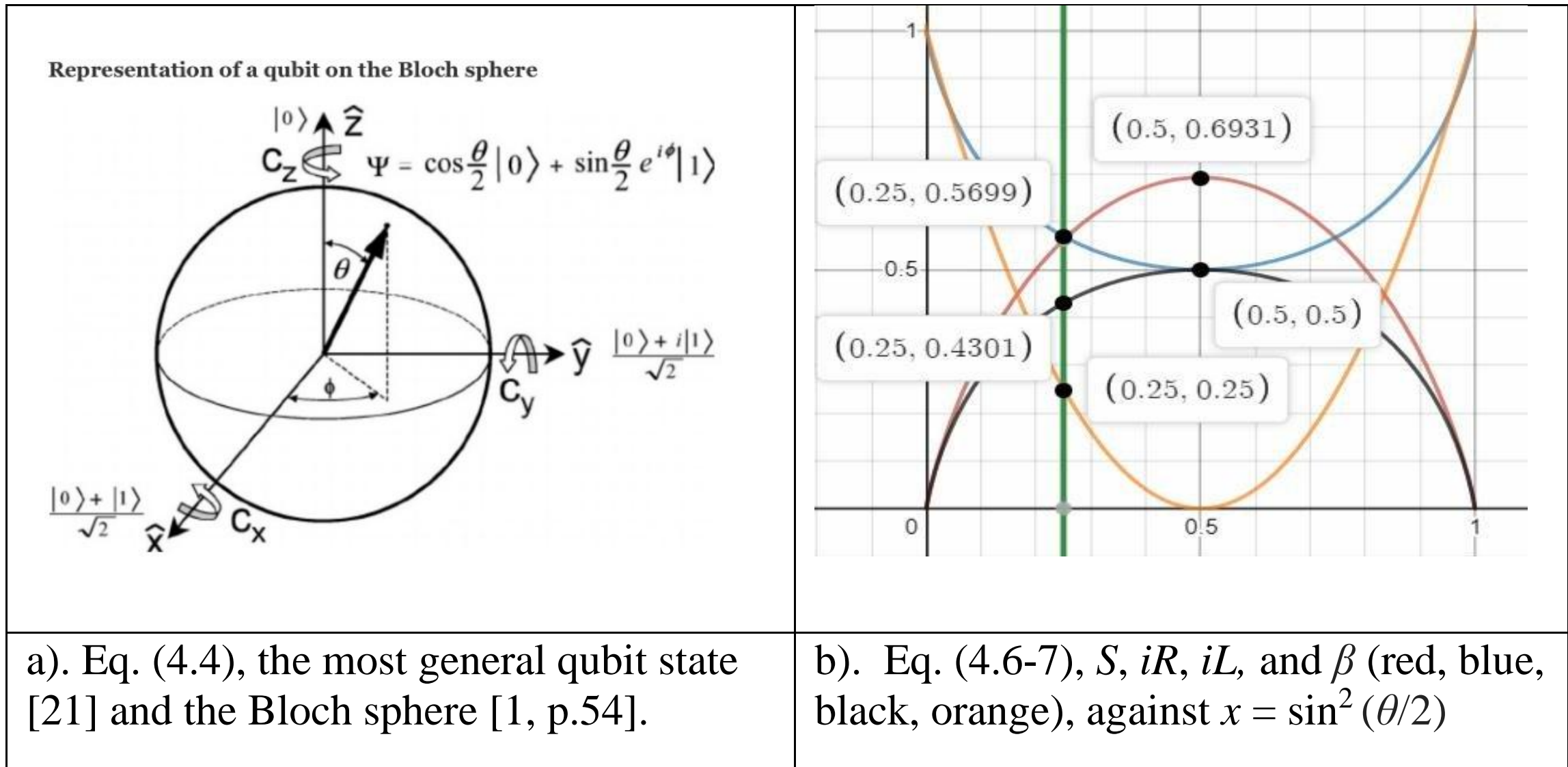

a). Eq. (4.4), the most general qubit state [21] and the Bloch sphere [1, p.54].

b). Eq. (4.6-7), $S$, $iR$, $iL$, and $\beta$ (red, blue, black, orange), against $x = \sin^2(\theta/2)$

Fig. 4.1: The general 1-q state and its QE, IR, IL, and $\beta$ ([22, Desmos4.1])[2]

---

[2] We use the online graphic calculator of Desmos.com [22] to draw and store the curves of entropy, information retrievability, and loss for different quantum systems.

As shown in Fig (4.1b), at $\theta = 0$ (|q⟩ = |0⟩), or $\theta = \pi$ (|q⟩ =|1⟩), it has the maximal bias ($\beta$ =1), zero entropy and zero info loss. At $\theta = \pi/2$, the state is |q⟩ = (1/√2) (|0⟩+$e^{i\varphi}$|1⟩, as in Eq. (2.8), it has zero bias ($\beta = 0$), maximal entropy ln 2, and maximal info loss $e^{-\ln 2} = 1/2$. Meanwhile, the azimuthal angle $\varphi$, related to quantum interference [14], is erased by measurements.

Table 4.1 lists three characteristic bias cases (from the maximum to minimum).

| Polar angle $\theta$ | 0, π | π/3 | π/2 |
|---|---|---|---|
| $x = \sin^2(\theta/2)$ | 0 | 0.25 | 0.5 |
| Polar bias (orange): $\beta$ ($\theta$) | 1 (maximal) | 1/4 = 0.25 | 0 (unbiased) |
| Entropy (red): $S(\theta)$ | 0 | 0.5623 | ln 2 = 0.6931 |
| Retrievable info (blue): $iR(\theta)$ | 1 | 0.5699 | ½ = 0.5 |
| Lost info (black): $iL(\theta)$ | 0 | 0.4301 | ½ = 0.5 |

Table 4.1: The dependence of QE, IR, IL, and bias $\beta$ on $\theta$

At $\theta = \pi/2$ (no bias), the state has maximal entropy $S_2(\rho_{\max})$, and, from (4.5), it is an MMS with a PMF of discrete uniform distribution [2, p. 11], described in Eq. (3. 18-21) with $m$ = 1:

$$P(0\,|\,\Omega_{\rho_{\max}}) = P(1\,|\,\Omega_{\rho_{\max}}) = 1/2, \quad \Omega_{\rho_{\max}} = \{0,1\} \tag{4.9}$$

Such a PMF and its 1-bit classical entropy are identical to tossing one fair coin without any quantum information left.

## 5. Bell States, Entanglement Entropy and Mutual Quantum Entropy

The Bell test uses Bell states to verify Bell inequalities [4, p.115]. Two entangled photons are involved, and we need two observers (Alice and Bob) who might be located a light-year away from each other.

The device setup is described in Fig. 5.1. The photons move in the $x$-direction left to Alice (a) and right to Bob (b). The two Electro-Optic Modulators (EOM) can be rotated in the $y$-$z$ plane (with angle {$\theta$, 0, −θ} respectively, perpendicular to the paper) to change the polarization direction of the photons. Then, a polarizing beam splitter (PBS or Polarizer) at each side splits the photon, which will either go to the vertical detector (Red light) or horizontal detector (Green light).

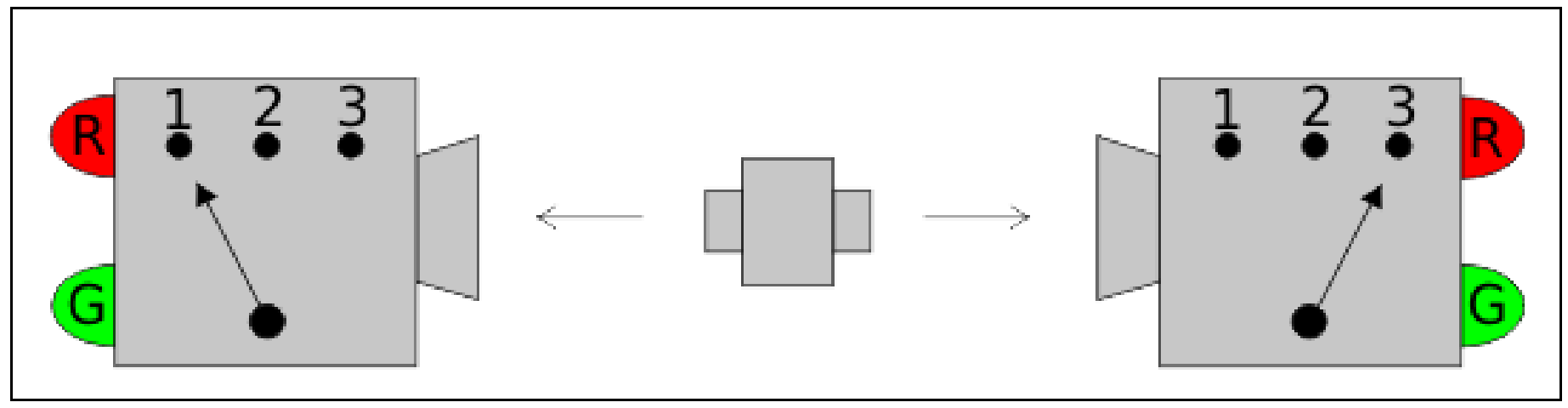

Fig. 5.1: The simplified setup of the Bell test (from the web)

When the EOM is at position 2, it is in the $z$-direction (up-down), and the R/G detected signals are denoted by $V$ (in the $z$-direction) and $H$ (in the $y$-direction). At position 1 (3), the EOM has an angle of $\theta$ ($-\theta$) with a $z$-axis, and the detected signals are denoted by (V′, H′) and (V″, H″), respectively. The relative polarization angles between the EOM settings are denoted by $\theta_{a,b}$, ∈ {0, $\theta$, 2$\theta$}. The directions of the two EOMs are independently chosen at random.

At position 2, the polarization is in the $z$-direction; the four computational basis states (CB) are:

$$|HH\rangle \equiv |H\rangle_A H\rangle_B,\ |VV\rangle,\ |HV\rangle,\ |VH\rangle \tag{5.1}$$

There are four Bell states [4, p. 25] orthogonal to each other:

$$|\Phi^{\pm}\rangle = \frac{1}{\sqrt{2}}(|HH\rangle \pm |VV\rangle) \tag{5.2}$$

$$|\Psi^{\pm}\rangle = \frac{1}{\sqrt{2}}(|HV\rangle \pm |VH\rangle) \tag{5.3}$$

Because they are all pure states, they have zero joint quantum entropy:

$$S(\rho_{AB}) = S(|\Phi^{\pm}\rangle\langle\Phi^{\pm}|\ ) = S(|\Psi^{\pm}\rangle\langle\Psi^{\pm}|\ ) = 0 \tag{5.4}$$

Bell states are maximally entangled [1, p.98] since their reduced density matrix of each 2D subsystem (A or B) has the maximum entropy ($S$ (ρ) = ln $N$ = ln 2):

$$\rho_{AB} = |\Phi^{\pm}\rangle\langle\Phi^{\pm}| = \frac{1}{2}\left(|H\rangle_A|H\rangle_B \pm |V_A\rangle|V_B\rangle\right)\left({}_A\langle H|\,{}_B\langle H| \pm {}_A\langle V|\,{}_B\langle V|\right)$$

$$\rho_A = \mathrm{Tr}_B\rho_{AB} = \frac{1}{2}\sum_{x=\in\{H,V\}} {}_B\langle x|\left(|H\rangle_A|H\rangle_B \pm |V_A\rangle|V_B\rangle\right)\left({}_A\langle H|\,{}_B\langle H| \pm {}_A\langle V|\,{}_B\langle V|\right)|x\rangle_B$$

$$= \frac{1}{2}\left(|H\rangle_A\,{}_A\langle H|)(|V_A\rangle_A\langle V|\right) = \frac{1}{2}\begin{pmatrix}1 & 0\\ 0 & 1\end{pmatrix} \tag{5.5}$$

$$\therefore\ S(\rho_A) = -\mathrm{Tr}\,(\rho_A \ln \rho_A) = -\sum_{\alpha=1}^{2}\frac{1}{2}\ln\frac{1}{2} = \ln 2 \triangleq EE_A \tag{5.6}$$

Similarly: $$S(\rho_B) = -\mathrm{Tr}\,(\rho_B \ln \rho_B) = -\sum_{\alpha=1}^{2}\frac{1}{2}\ln\frac{1}{2} = \ln 2 \triangleq EE_B \tag{5.7}$$

Here, EE stands for Entanglement Entropy [23]. The total entropy in the direct product of the density matrices is given by (see App A):

$$S_{A\otimes B} = S(\rho_A \otimes \rho_B) = S(\rho_A) + S(\rho_B) = 2\ln 2 = \ln 4 \tag{5.8}$$

It contains the same information as $\rho_A$ and $\rho_B$ contain together: $\rho_{A\otimes B}$ describes a statistical ensemble for which the variables of subsystem $A$ are completely uncorrelated with the variables of subsystem $B$. Assuming the initial state is the Bell state $\rho_{AB}$, and the final state is $\rho_{A\otimes B}$, we have the following change of entropy:

$$S(\rho_{A\otimes B}) - S(\rho_{AB}) = S(\rho_A) + S(\rho_B) - S(\rho_{AB}) \triangleq I(A:B) \triangleq MQE_{A:B} \qquad (5.9$$

We refer $I(A{:}\,B)$ to the mutual quantum entropy (MQE)[3]. From Eq. (3.13), the comparative retrievability reads:

$$iR(S_{A\otimes B}, S_{AB}) = e^{S(\rho_{AB})-S(\rho_A\otimes\rho_B)} \triangleq e^{-I(A:B)} = iR(I(A:B)) \triangleq iR(MQE_{A:B}) \tag{5.10}$$

For Bell states, $S(\rho_{AB}) = 0$, we have

$$MQE_{Bell} = S_{A\otimes B} = \ln 4, \qquad iR(MQE_{Bell}) = e^{-\ln 4} = 1/4 = 25\% \tag{5.11}$$

The Hilbert space has dimension $N_{AB} = 4$, so $MQE_{Bell}$ is maximal, representing an MMS with $m$ = 2, indicating the total loss of quantum information due to the gain of $MQE_{Bell}$:

$$iL_{Bell} \triangleq iL(MQE_{Bell}) = 1 - iR(MQE_{Bell}) = 3/4 = 75\% \tag{5.12}$$

The density matrix of the uncorrelated subsystems can be written by using Eq. (5.5-5.7):

$$\begin{aligned}\rho(A\otimes B) &= \frac{1}{2}(|H\rangle\langle H| + |V\rangle\langle V\rangle)_A \otimes \frac{1}{2}(|H\rangle\langle H| + |V\rangle\langle V|)_B \\ &= 1/4\ (|HH\rangle\langle VV| + |HV\rangle\langle HV| + |VH\rangle\langle VH| + |VV\rangle\langle VV|)\end{aligned} \tag{5.13}$$

This is a 2-q MMS. Note that for Bell states, its MQE is the sum of the two EE's:

$$MQE_{Bell} = EE_A + EE_B = 2\ln 2,\ iL_{Bell} = 1 - iR(EE_A)\cdot iR(EE_B) = 3/4 \tag{5.14}$$

However, measurements may not cause information loss. If the initial state is a 2-qubit separable pure state $|HH\rangle = |H\rangle_A \otimes |H\rangle_B$ [1, p.155], where each photon is maximally biased as in Eq. (2.4-6), and if both Alice and Bob take the position (2, 2) = (0°, 0°), then the state is unchanged by measurements in the CB of Eq. (5.1). There is neither info loss nor entropy gain:

$$\begin{aligned}&P(HH\,|\,\Omega_{HH}) = |\langle HH\,|\,HH\rangle|^2 = 1, \quad P(HV\,|\,\Omega_{HH}) = P(HV\,|\,\Omega_{HH}) = P(VV\,|\,\Omega_{HH}) = 0 \\ &\therefore\ \rho_{HH,i} = \rho_{HH,f} = |HH\rangle\langle HH|, \quad S(\rho_{HH,i}) = S(\rho_{HH,f}) = 0 \\ &\therefore\ qR(S_f, S_i) = e^{S_i - S_f} = 1, \quad qL(S_f, S_i) = 0\end{aligned} \tag{5.15}$$

## 6. Information Loss of a Bell State in Bell Tests

Now, let us assume that Alice and Bob share a Bell state (it can be any of the 4 Bell states):

$$|\Phi^+\rangle = \frac{1}{\sqrt{2}}(|H\,H\rangle + |V\,V\rangle) \tag{6.1}$$

We investigate two specific measurement settings.

[3] We prefer to call I (A: B) the mutual quantum entropy, as it is entropy by definition, and its gain indicates the loss of related mutual information. Of course, it can also be referred to as quantum mutual information.

**Case 1**. Position (a, b) = (2, 2) = (0°, 0°), $\theta_{a,b} = 0$**:** Both EOM are at position 2, no rotation. There are four possible outcomes (CB vectors) from Alice and Bob, as in Eq. (5.1). It is trivial to find their joint probability mass function:

$$P(HH \mid \Omega_{\Phi^+}) = |\langle HH \mid \Phi^+ \rangle|^2 = \frac{1}{2} = P(VV \mid \Omega_{\Phi^+}), \quad P(HV \mid \Omega_{\Phi^+}) = P(HV \mid \Omega_{\Phi^+}) = 0 \quad (6.2)$$

The density matrix corresponding to the PMF is:

$$\rho_{zz} = |\,H\,H\,\rangle \frac{1}{2} \langle H\,H\,| + |\,V\,V \rangle \frac{1}{2} \langle V\,V\,| \quad (6.3)$$

The entropy, retrievable, and lost quantum info reads:

$$S(\rho_{zz}) = -\mathrm{Tr}\,(\rho_{zz} \ln \rho_{zz}) = -\sum_{\alpha=1}^{2} \frac{1}{2} \ln \frac{1}{2} = \ln 2 \quad (6.4)$$

$$iR(\rho_{zz}) = e^{-\ln 2} = 0.5, \quad iL(\rho_{zz}) = 0.5 \quad (6.5)$$

Consequently, if Alice and Bob measure {H, V} and share their results through classical communication, the expected information loss is 50%.

Now, suppose that Alice does not tell Bob what she sees. What can Bob do with his observations? From the joint PMF (6.2), he can only find:

$$P(H_B \mid \Omega_{\Phi^+}) = \sum_{\chi_A \in \{H,V\}} P(\chi_A, H_B \mid \Omega_{\Phi^+}) = \frac{1}{2}, \; P(V_B \mid \Omega_{\Phi^+}) = \sum_{\chi_A \in \{H,V\}} P(\chi_A, V_B \mid \Omega_{\Phi^+}) = \frac{1}{2} \quad (6.6)$$

To Bob, the state of his photon now is an MMS, similar to Eq. (5.7):

$$\rho_B = |\,H\rangle_B \frac{1}{2}\, {}_B\langle H\,| + |\,V\rangle_B \frac{1}{2}\, {}_B\langle V\,|, \quad S(\rho_B) = \ln 2, \quad iR(\rho_B) = \frac{1}{2} \quad (6.7)$$

So, by examining his data, Bob cannot determine if his photon was previously entangled with another photon. The same is true for Alice. The total entropy, now increased by ln2, is equal to $MQE_{Bell}$, corresponding to information loss $iL_{Bell}$, as given in Eq. (5.12):

$$S = S(\rho_{zz}) + \ln 2 = \ln 4 = MQE_{Bell}, \; iR(MQE_{Bell}) = e^{-2\ln 2} = 1/4, \; iL(MQE_{Bell}) = 3/4 \quad (6.8)$$

Hence, if Alice and Bob don't communicate classically, the system gains extra entropy and loses extra information (more on this later in this section). But if they exchange records later, even years later, they can recover this lost information and remove the corresponding entropy.

**Case 2**. Position (a, b) = (1, 3) = ($\theta$, $-\theta$) or $\theta_{a,b} = 2\theta$: The HWP at Alice's site is rotated by $\theta$ (position 1), at Bob's by $-\theta$ (position 3). The four possible outcomes from Alice and Bob are: *H′H″*, *H′V″*, *V′H″*, and *V′V″*. The rotated basis can be represented as (see the Bloch sphere in Fig. 4.1a):

$$|\,H'\rangle = \cos\left(\frac{\theta}{2}\right)|\,H\rangle - \sin\left(\frac{\theta}{2}\right)|V\rangle, \quad |V'\rangle = \sin\left(\frac{\theta}{2}\right)|\,H\rangle + \cos\left(\frac{\theta}{2}\right)|V\rangle \quad (6.9)$$

$$|H''\rangle = \cos\left(\frac{\theta}{2}\right)|H\rangle + \sin\left(\frac{\theta}{2}\right)|V\rangle \quad |V''\rangle = -\sin\left(\frac{\theta}{2}\right)|H\rangle + \cos\left(\frac{\theta}{2}\right)|V\rangle \tag{6.10}$$

From Eq. (6.1), the PMF can be calculated easily:

$$\begin{aligned} P(V',V'') &= P(V',V''|\Omega_{\Phi^+}) = \left|\langle\Phi^+|V'\rangle_A|V''\rangle_B\right|^2 \\ &= \frac{1}{2}\left(-\sin^2\left(\frac{\theta}{2}\right)+\cos^2\left(\frac{\theta}{2}\right)\right)^2 = \frac{1}{2}\cos^2\theta = P(H',H'') \end{aligned} \tag{6.11}$$

$$\begin{aligned} P(V',H'') &= P(V',H''|\Omega_{\Phi^+}) = \left|\langle\Phi^+|V'\rangle_A|H''\rangle_B\right|^2 \\ &= \frac{1}{2}\left(2\sin\left(\frac{\theta}{2}\right)\cos\left(\frac{\theta}{2}\right)\right)^2 = \frac{1}{2}\sin^2\theta = P(H',V'') \end{aligned} \tag{6.12}$$

Similarly, for position (a, b) = (1, 2) = ($\theta$, 0) or $\theta_{a,b} = \theta$, one can find

$$\begin{aligned} P(H,H') &= P(H,H'|\Omega_{\Phi^+}) = \left|\langle\Phi^+|H\rangle_A|H'\rangle_B\right|^2 \\ &= \frac{1}{2}\left|\langle H|\left(\cos\left(\frac{\theta}{2}\right)|H\rangle - \sin\left(\frac{\theta}{2}\right)|V\rangle_B\right)\right|^2 = \frac{1}{2}\cos^2\left(\frac{\theta}{2}\right) = P(V,V') \end{aligned} \tag{6.13}$$

$$\begin{aligned} P(H,V') &= P(H,V'|\Omega_{\Phi^+}) = \left|\langle\Phi^+|H\rangle_A|V'\rangle_B\right|^2 \\ &= \frac{1}{2}\left|\langle H|\left(\sin\left(\frac{\theta}{2}\right)|H\rangle + \cos\left(\frac{\theta}{2}\right)|V\rangle_B\right)\right|^2 = \frac{1}{2}\sin^2\left(\frac{\theta}{2}\right) = P(V,H') \end{aligned} \tag{6.14}$$

If we set $\theta \to \theta/2$ in PMF (6.11-12), that is, redefine $\theta_{a,b} = \theta$, we get the same PMF (6.13-14). This means that PMF's period in (6.11-12) is half that of PMF's period in (6.13-14).

From (6.13) and (6.14), for (a, b) = (1, 2) = ($\theta$, 0), we have the following density matrix:

$$\rho(\theta) = \frac{1}{2}\cos^2\left(\frac{\theta}{2}\right)\left(|HH'\rangle\langle HH'| + |VV'\rangle\langle VV'|\right) + \frac{1}{2}\sin^2\left(\frac{\theta}{2}\right)\left(|HV'\rangle\langle HV'| + |VH'\rangle\langle VH'|\right) \tag{6.15}$$

It leads to the following QE, IR, IL, and the mutual bias $\beta$, as shown in Fig (6.1):

$$S(\theta) = -\cos^2\left(\frac{\theta}{2}\right)\ln\left(\frac{1}{2}\cos^2\left(\frac{\theta}{2}\right)\right) - \sin^2\left(\frac{\theta}{2}\right)\ln\left(\frac{1}{2}\sin^2\left(\frac{\theta}{2}\right)\right) \quad \text{(red)} \tag{6.16}$$

$$iR(\theta) = \left(\frac{1}{2}\cos^2\left(\frac{\theta}{2}\right)\right)^{\cos^2\left(\frac{\theta}{2}\right)}\left(\frac{1}{2}\sin^2\left(\frac{\theta}{2}\right)\right)^{\sin^2\left(\frac{\theta}{2}\right)} \quad \text{(blue)} \tag{6.17}$$

$$iL(\theta) = 1 - iR(\theta) \quad \text{(black)} \tag{6.18}$$

$\beta(\theta) \triangleq [\cos^2(\theta/2) - \sin^2(\theta/2)]^2 = \cos^2(\theta)$, same as (4.8) (orange)

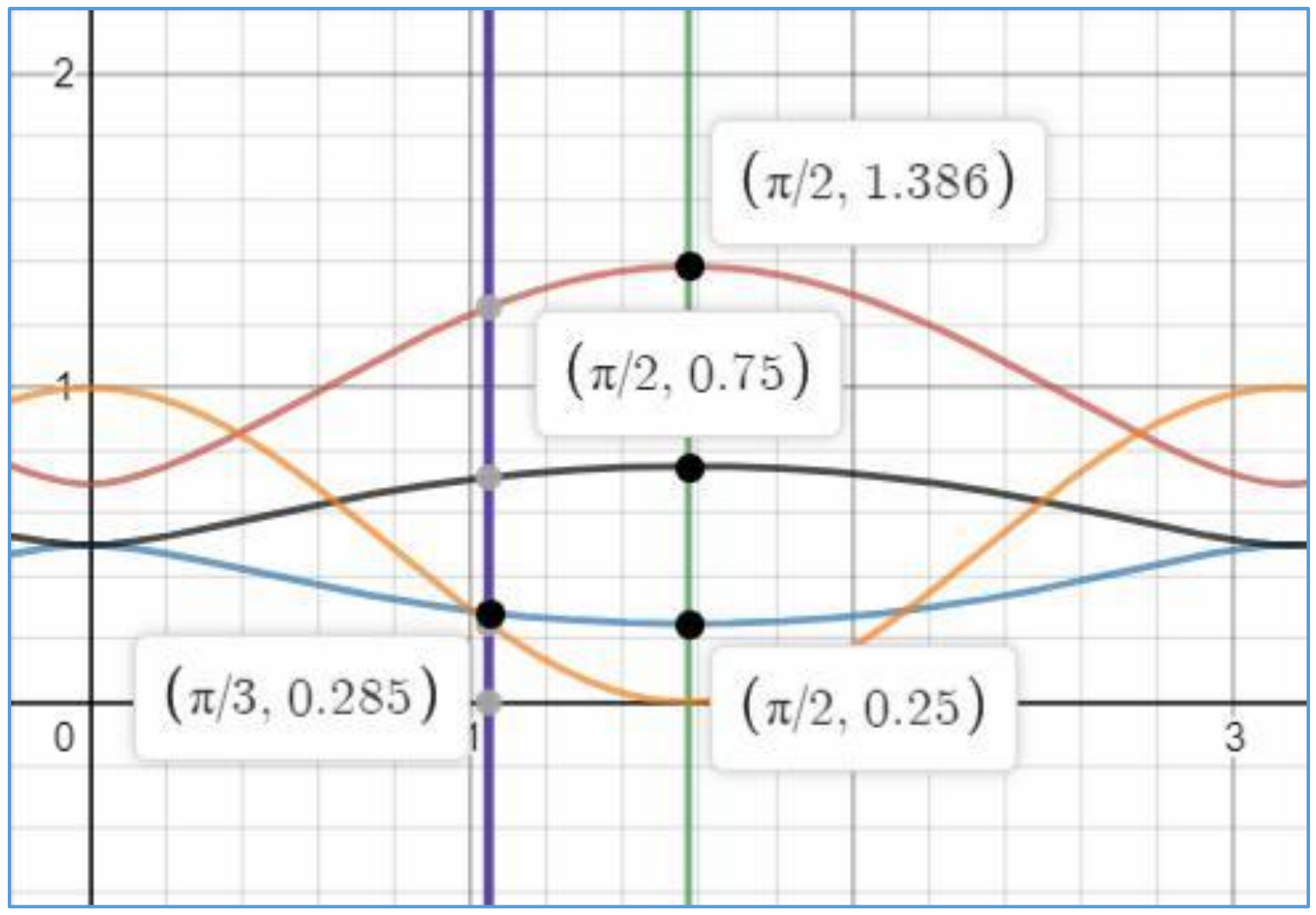

Fig. 6.1: QE, IR, IL, and $\beta$ (red, blue, black, orange) against $\theta$ ([22, DESMOS 6.1])

The IR and IL for the three representative relative polarization angles (0, π/3, π/2) are:

Retrievable (blue): $0.5 = qR(\rho_{\theta=0}^{\beta=1}) > qR(\rho_{\theta=\pi/3}^{\beta=0.25}) = 0.285 > qR(\rho_{\theta=\pi/2}^{\beta=0}) = 0.25$ (Min) (6.19)

Lost (black): $0.5 = qL(\rho_{\theta=0}) < qL(\rho_{\theta=\pi/3}) = 0.715 < qL(\rho_{\theta=\pi/2}) = 0.75$ (Max) (6.20)

| Polar angle $\theta$ | 0, π | π/3 | π/2 |
|---|---|---|---|
| Mutual bias (orange): $\beta$ ($\theta$) | 1 (maximal) | 1/4 = 0.25 | 0 (unbiased) |
| Entropy (red): $S(\theta)$ | ln 2 = 0.693 | 1.255 | ln 4 = 1.386 |
| Retrievable info (blue): $iR(\theta)$ | 0.5 | 0.285 | 0.25 |
| Lost info (black): $iL(\theta)$ | 0.5 | 0.715 | 0.75 |
| Removable entropy | ln 2 = 0.693 | 0.131 | 0 |

Table 6.1: The dependence of QE, IR, IL, and fairness $\beta$ on $\theta$

At $\theta = \pi/2$ (no bias), the state is an $m = 2$ MMS with the following basis vectors:

$$|00\rangle = |HH'\rangle,\ |01\rangle = |HV'\rangle,\ |10\rangle = |VH'\rangle,\ |11\rangle = |VH'\rangle \tag{6.21}$$

**Comments on Bell Tests:**

(**6A**): From Fig 6.1 or Eq. (6.22), at least 50% of info is lost by measuring a Bell state, while the minimal loss is 0% for the separable pure state [1, p.155] described in Eq. (5.15). This characterizes the unescapable loss of the *minimal entanglement information* (MEI) associated with the gain of the Minimal Entanglement Entropy (MEE) for a Bell state:

$$MEE\ (\text{Bell}) \triangleq \min S\ (\text{Bell}) = \ln 2, \quad \min iR\ (\text{Bell}) = e^{-\ln 2} = ½ \tag{6.22}$$

$$MEI\ (\text{Bell}) \triangleq \min iL\ (\text{Bell}) = 1 - \min iR\ (\text{Bell}) = 1 - e^{-\ln 2} = ½ \tag{6.23}$$

(**6B**): Alice and Bob must exchange records to verify Bell's inequality. Otherwise, from the joint PMF (6.11-12), Bob can only find his photon in an MMS with entropy ln 2:

$$P(H''_B \mid \Omega_{\Phi^+}) = \sum_{\chi_A \in \{H', V'\}} P(\chi_A, H''_B \mid \Omega_{\Phi^+}) = \frac{1}{2}, \;\; P(V''_B \mid \Omega_B) = \frac{1}{2} \tag{6.24}$$

The same is true for Alice. The total entropy is now maximal (2 ln2). As in Eq. (6.8) and (3.21), it has 2 bits of Shannon classical entropy for tossing two fair coins [1, p.128], losing all original MQI (interference, entanglement, and mutual bias). The extra lost info is the quantum info related to mutual bias angle $\theta$, associated with the following extra quantum entropy gain:

$$\Delta S\,(\theta) = 2\ln 2 - S\,(\theta) \geq 0 \tag{6.25}$$

where $S\,(\theta)$ is given by Eq. (6.16). When $\theta = 0$, it leads to Eq. (6.8), an extreme case:

$$\Delta S\,(0) = 2\ln 2 - S\,(0) = \ln 2 \text{ (the maximal extra quantum entropy gain)} \tag{6.26}$$

Moreover, without exchanging records, the two photons of Alice and Bob would behave like a separable pure state[1, p.155], each photon in an unbiased state as in Eq. (2.9):

$$|\Psi_{AB}\rangle = \frac{1}{2}(|H\rangle_A + e^{i\varphi_A}|V\rangle_A) \otimes (|H''\rangle_B + e^{i\varphi_B}|V''\rangle_B) \tag{6.27}$$

No correlation exists, and the Bell inequality [4, p.115] would never be violated [6]. However, if they store their data, they can exchange it afterward, even many years later, thus confirming the violation of the Bell inequality, recovering the extra lost information, removing the extra entropy gain, and decreasing the total entropy to $S(\theta)$ in Eq. (6.16).

(**6C**): How can entropy decrease years later? The dilemma reminds us of the recoverable info in delayed choice [6,7]. In the experiment conducted by Kim et al. [6], Bob needs to obtain Alice's data regarding the path taken by the idler photon—whether it was one way or both ways—to correctly sort his already registered data for the signal photon. If he does not get this information, the details about quantum interference will be lost, and extra quantum entropy will be gained.

Similarly, in the experimental delayed-choice entanglement swapping by Ma et al. [7], Alice and Bob need to know Victor's choice and his results to sort their previously recorded data. Without this information, the quantum details of their two photons, whether entangled or separable, will be lost, which means there is additional quantum entropy gain.

Therefore, by sharing recorded knowledge in such cases, "the intervention of intelligent beings" (Szilard 1929 [24]) could remove extra quantum entropy gain years later. This may offer a new way to explain the EPR paradox [25] in Bell tests: without removing such extra quantum entropy gain, the Bell inequality could not be validated (or not be violated [26]), and there is no "spooky action"; after removing it, the Bell inequality can be violated, but data exchange can't go faster than light, so there is no "spooky action at a distance" either.

(**6D**): **The entropy uncertainty** [11-12] in the Bell test can be found using Eq. (6.9-10):

$$H(A)+H(B)\ge -\log c,\quad c=\max_{i,j}|\langle B_i \mid A_j\rangle|^2=\max[\cos^2(\theta/2),\sin^2(\theta/2)]$$

$$H(A)+H(B)\ge 0,\ \text{if } \theta=0 \text{ or } \pi;\ H(X)+H(K)\ge \ln 2,\ \text{if } \theta=\pi/2$$

It is consistent with our results in Table 6.1. When θ equals zero or π, the polarization direction of Alice and Bob are parallel, and the two observables are commuting and have no uncertainty. When θ is π/2, there is the maximum uncertainty, and the total VNE is maximal.

## 7. Quantum Teleportation and Biseparable Tripartite Systems

The biseparable states of three qubits are widely used in quantum teleportation [4, p.26] [9].

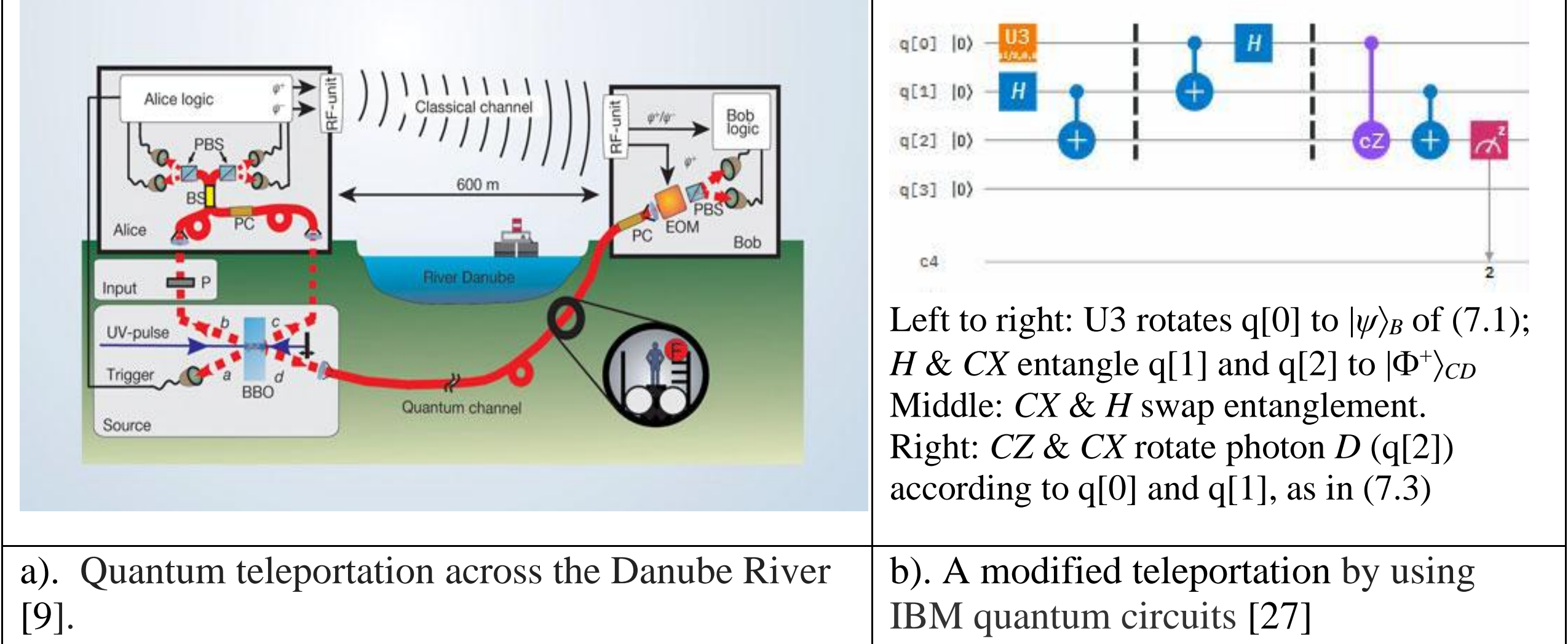

Left to right: U3 rotates q[0] to $|\psi\rangle_B$ of (7.1); *H* & *CX* entangle q[1] and q[2] to $|\Phi^+\rangle_{CD}$
Middle: *CX* & *H* swap entanglement.
Right: *CZ* & *CX* rotate photon *D* (q[2]) according to q[0] and q[1], as in (7.3)

a). Quantum teleportation across the Danube River [9].

b). A modified teleportation by using IBM quantum circuits [27]

Fig. 7.1: Quantum teleportation: an experiment and a quantum circuit

Assume that Alice and Bob share a pair of photons *C* and *D* (*c* and *d* in Fig. 7.1a), entangled in Bell state $|\Phi^+\rangle_{CD}$, and Alice has her own photon *B* rotated to the following state (Fig. 7.1b):

$$|\psi\rangle_B=\alpha\,|0\rangle_B+\beta\,|1\rangle_B\,,\ |\alpha|^2+|\beta|^2=1 \tag{7.1}$$

This state is to be teleported. Before swapping, Alice has two photons (*B* and *C*, as b and c in Fig. 7.1), while Bob has one photon (D, as *d* in Fig. 7.1). The state of the three photons is given by a separable pure state in the product Hilbert space $\mathcal{H}^8$:

$$|\Phi^+\rangle_{CD}\otimes|\psi\rangle_B=\frac{1}{\sqrt{2}}\big(|0\rangle_C\otimes|0\rangle_D+|1\rangle_C\otimes|1\rangle_D\big)\otimes\big(\alpha\,|0\rangle_B+\beta\,|1\rangle_B\big) \tag{7.2}$$

Then, photons *C* and *B* are transformed to the four Bell states by unitary transformation, forcing photon *D* to carry correspondingly rotated states from photon *B* [1. P.165]:

$$\begin{aligned}|\Phi^+\rangle_{CD}\otimes|\psi\rangle_B=\frac{1}{2}\Big[&|\Phi^+\rangle_{CB}\otimes(\alpha\,|0\rangle_D+\beta\,|1\rangle_D)+|\Phi^-\rangle_{CB}\otimes(\alpha\,|0\rangle_D-\beta\,|1\rangle_D)\\&+|\Psi^+\rangle_{CB}\otimes(\beta\,|0\rangle_D+\alpha\,|1\rangle_D)+|\Psi^-\rangle_{CB}\otimes(\beta\,|0\rangle_D-\alpha\,|1\rangle_D)\Big]\end{aligned} \tag{7.3}$$

Because the transformation is unitary at this time, it is still a pure state, and the information is conserved. Then Alice uses two sets of devices to measure photons *C* and *B*, deciding which Bell state they are in. Based on the result, Alice sends Bob two bits of information in a classical way, telling him how to rotate the state of photon *D* [9].

Now, what are the QE and IR after Alice's measurements? Alice, only acting on photons *C* and *B*, gets the following density matrix in the 4-d Hilbert space $\mathcal{H}^4$ for her part:

$$\rho_{CB,Bell} = \frac{1}{4}\left[ |\Phi^+\rangle_{CB}\langle\Phi^+| + |\Phi^-\rangle_{CB}\langle\Phi^-| + |\Psi^+\rangle_{CB}\langle\Psi^+| + |\Psi^-\rangle_{CB}\langle\Psi^-| \right] \qquad (7.4)$$

Since the 4 Bell states form a complete set of orthogonal vectors, density matrix $\rho_{CB,Bell}$ describes a maximal mixed state (MMS) for $m = 2$ as in Eq. (3.17):

$$S(\rho_{CB}) = -\sum_{i=1}^{4}\frac{1}{4}\ln\frac{1}{4} = \ln 4 = 2\ln 2, \quad iR(\Delta S_{CB}) = \frac{1}{4}, \quad iL = \frac{3}{4} \qquad (7.5)$$

Hence, the entropy gain and the information loss are both maximal. Alice will obtain two bits of classical information for each measurement, describing which Bell state she gets.

After Alice's measurement, photon D becomes one of the four pure states, as shown in Eq. (7.3). Then, Bob, using Alice's 2 bits message, rotates (a unitary transformation!) his photon *D* to recover the state in Eq. (7.1). Without doing any measurements, Bob's qubit is still a pure stat; hence, the system's total entropy, IR and IL are the same as in Eq. (7.5).

Now assume that Alice fails to send Bob her data promptly, so Bob does not know how to rotate his photon correctly in time. Can he use Alice's record later? Unfortunately, it might be a mission impossible. Unlike Allice, who can easily record the 2-bit classical info for each photon and keep them forever, Bob's hands are tied theoretically by the "No-cloning theorem" [3, p.24, p.532] and practically by the limitation of current quantum memory in size and stability (timeline of quantum computing [26]). Once again, we see the vital importance of classical communication in bipartite quantum experiments.

**The entropy uncertainty in the teleportation** [11-12]: From the definition of Bell states,

$$H(X) + H(K) \geq -\log c, \quad c = \max_{i,j} |\langle\Phi_i | \Psi_j\rangle|^2 = 1/2$$
$$H(X) + H(K) \geq \ln 2$$

Alice's measurement results in an entropy of 2 ln2, greater than the lower uncertainty bound of ln2. Bob does not measure photon D and does not generate any entropy.

## 8. Measurements of Three or More Entangled Qubits

There are two important non-biseparable classes of 3-q entanglement states [10, 29].

**Class 1**. *A GHZ* (Greenberger-Horne-Zeilinger) state [10, 29] of three qubits is given by

$$|GHZ_3\rangle = \frac{1}{\sqrt{2}}\left(|000\rangle + |111\rangle\right) \tag{8.1}$$

It is a pure state, having zero QE. The Hilbert space is 8-dimensional. After measuring one qubit in CB, its reduced density matrix, entropy, and IR can be found as:

$$\rho_{GHZ,12} \equiv \mathrm{Tr}_3 \rho_{GHZ} = \frac{1}{2}\mathrm{Tr}_3(|000\rangle + |111\rangle)(\langle 000| + |\langle 111|) = \frac{1}{2}(|00\rangle\langle 00| + |11\rangle\langle 11|) \tag{8.2}$$

$$S(\rho_{GHZ,12}) = \ln 2 = 0.6931, \quad iR(S_{GHZ,12}) = e^{-\ln 2} = 1/2 = iL(S_{GHZ,12}) \tag{8.3}$$

**Class 2**. A *W state* [10,29] of three qubits is given by

$$|W_3\rangle = \frac{1}{\sqrt{3}}(|001\rangle + |010\rangle + |100\rangle) \tag{8.4}$$

It is also a pure state. After measuring one qubit in CB, its reduced density matrix reads:

$$\begin{aligned}
&\rho_{W,12} \equiv \mathrm{Tr}_3 \rho_W = ⅓\, \mathrm{Tr}_3(|001\rangle + |010\rangle + |100\rangle)(\langle 001| + \langle 010| + \langle 100|) \\
&= ⅓\,(|00\rangle\langle 00| + |01\rangle\langle 01| + |01\rangle\langle 10| + |10\rangle\langle 01| + |10\rangle\langle 10|) \\
&= ⅓\left(|01\rangle\langle 01| + |01\rangle\langle 10| + |10\rangle\langle 01| + |10\rangle\langle 10|\right) + ⅓\,|00\rangle\langle 00| \\
&= |\Psi^+\rangle\frac{2}{3}\langle\Psi^+| + |00\rangle\frac{1}{3}\langle 00|
\end{aligned} \tag{8.5}$$

Here $|\Psi^+\rangle$ is one of the four maximally entangled Bell states, as shown in Eq. (5.3):

$$|\Psi^+\rangle = \frac{1}{\sqrt{2}}(|01\rangle + |10\rangle), \quad \langle 00|\Psi^+\rangle = 0 \tag{8.6}$$

The QE and IR of the mixed state in Eq. (8.5) is given by:

$$S(\rho_{W,12}) = -\frac{2}{3}\ln\frac{2}{3} - \frac{1}{3}\ln\frac{1}{3} = \ln\left(\frac{3}{2^{2/3}}\right) = 0.6365 < \ln 2 = 0.6931 = S(\rho_{GHZ,12}) \tag{8.7}$$

$$iR(S_{W,12}) = 2^{2/3}/3 = 0.5291, \quad iL(S_{W,12}) = 0.4709 < 0.5 = iL(S_{GHZ,12}) \tag{8.8}$$

Thus, after a measurement, $W_3$ contains more information than $GHZ_3$. Besides, if we measure one of the subsystems of the $GHZ_3$ in such a way that the measurement distinguishes between the states 0 and 1, we will get either $|00\rangle$ or $|11\rangle$, both are separable pure states:

$$_3\langle 0|GHZ_3\rangle = \frac{1}{\sqrt{2}}|0\rangle_1 \otimes |0\rangle_2, \quad _3\langle 1|GHZ_3\rangle = \frac{1}{\sqrt{2}}|1\rangle_1 \otimes |1\rangle_2 \tag{8.9}$$

Contrasting to the $GHZ_3$ state, the $W_3$ state has a chance to keep a maximally entangled pair if we measure one of its subsystems [30] and distinguish the resulting states:

$$ {}_3\langle 0|W_3\rangle = \frac{1}{\sqrt{3}}(|01\rangle + |10\rangle) = \sqrt{\frac{2}{3}}\,|\Psi^+\rangle, \qquad {}_3\langle 1|W_3\rangle = \frac{1}{\sqrt{3}}|0\rangle_1 \otimes |0\rangle_2 \tag{8.10} $$

After the one-qubit measurement, $W_3$ can have a 2/3 chance to be an entangled pair, while $GHZ_3$ can have only separable pairs; thus $W_3$ state will have less entropy gain, higher retrievability, and less loss of entanglement information as shown in Eq. (8.7-8). This explains why $W_3$ states are more robust than the $GHZ_3$ states. By the way, both 3-q states can be created and tested by using IBM quantum circuits [27], as shown in Fig. (8.1.a) and (8.1.b)

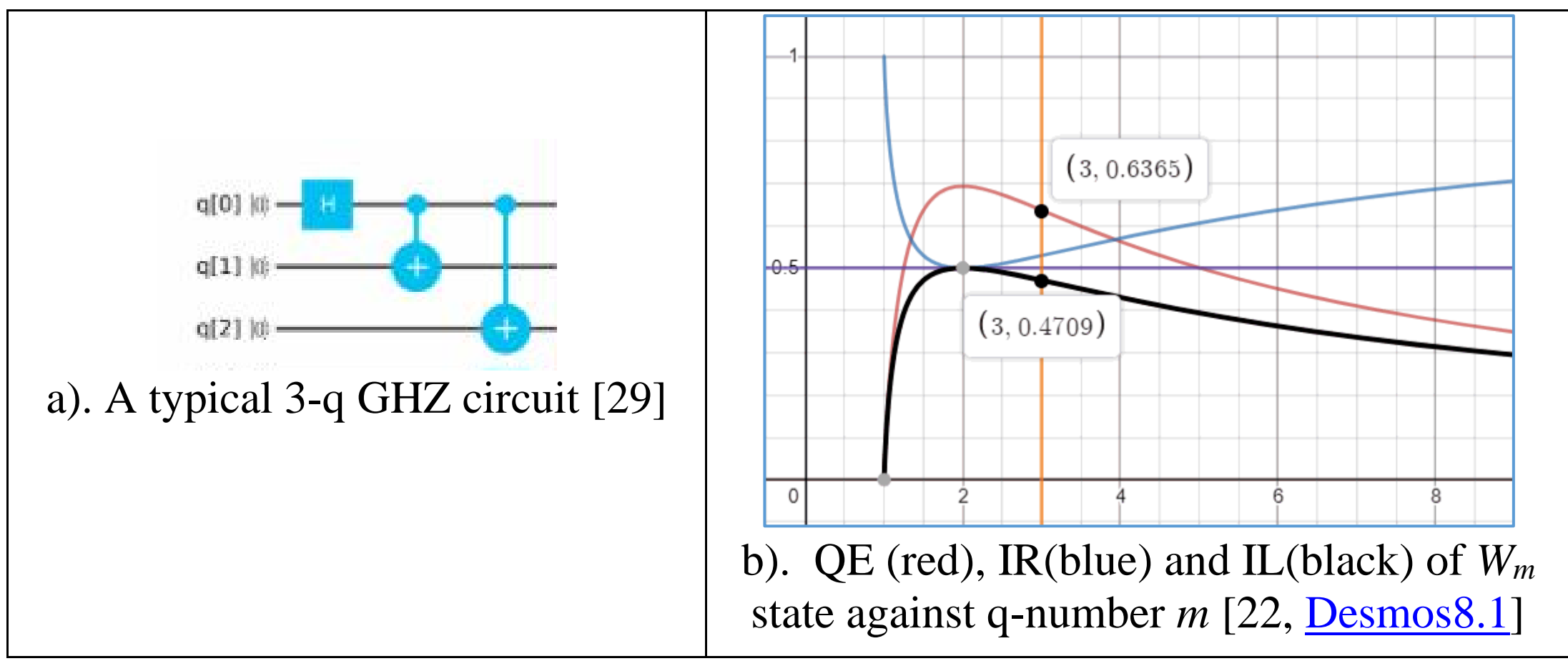

a). A typical 3-q GHZ circuit [29]

b). QE (red), IR(blue) and IL(black) of $W_m$ state against q-number $m$ [22, Desmos8.1]

Fig 8.1: ILC for GHZ$_3$ and W$_3$ states; QE, IR and IL of $W_m$ states

The *GHZ* and *W* states can be easily extended to systems of $m$ entangled particles ($m > 2$):

$$ |GHZ_m\rangle = \frac{1}{\sqrt{2}}\left(|0\rangle^{\otimes m} + |1\rangle^{\otimes m}\right) \tag{8.11} $$

$$ |W_m\rangle = \frac{1}{\sqrt{m}}(|100...0\rangle + |010...0\rangle + |00...01\rangle) \tag{8.12} $$

Similar to the 3-qubits cases, after measuring the first qubit in CB, we have for *GHZ* state:

$$ {}_1\langle 0|GHZ_m\rangle = \frac{1}{\sqrt{2}}|0\rangle^{\otimes(m-1)}, \quad {}_1\langle 1|GHZ_M\rangle = \frac{1}{\sqrt{2}}|1\rangle^{\otimes(m-1)} $$

The corresponding QE, IR, and IL are independent of $m$:

$$ \rho_{GHZ,m-1} \equiv \mathrm{Tr}_1 \rho_{GHZ,m} = \frac{1}{2}|0\rangle^{\otimes(m-1)}\langle 0| + \frac{1}{2}|1\rangle^{\otimes(m-1)}\langle 1| \tag{8.14} $$

$$ S(\rho_{GHZ,m-1}) = \ln 2 = 0.6931, \quad iR(S_{GHZ,m-1}) = e^{-\ln 2} = 1/2 = iL(S_{GHZ,m-1}) \tag{8.15} $$

As for the *W* state, it has a chance to keep entanglements after measuring the first qubit:

$$ {}_1\langle 0|W_m\rangle = \frac{\sqrt{m-1}}{\sqrt{m}}|W_{m-1}\rangle, \qquad {}_1\langle 1|W_m\rangle = \frac{1}{\sqrt{m}}|0\rangle^{\otimes(m-1)} \tag{8.16} $$

The corresponding density matrix, QE, IR, and IL are ($m > 2$):

$$\rho_{W,m-1} \equiv \mathrm{Tr}_1 \rho_{W,m} = \frac{m-1}{m} | W_{m-1} \rangle^{\otimes(m-1)} \langle W_{m-1} | + \frac{1}{m} | 0 \rangle^{\otimes(m-1)} \langle 0 | \quad (8.17)$$

$$S(\rho_{W,m-1}) = -\frac{m-1}{m} \ln \frac{m-1}{m} - \frac{1}{m} \ln \frac{1}{m} = \ln \frac{m}{(m-1)^{1-1/m}} \quad (8.18)$$

$$iR(S_{W,m-1}) = e^{-S} = \frac{1}{m}(m-1)^{1-1/m}, \quad iL(S_{W,m-1}) = 1 - \frac{1}{m}(m-1)^{1-1/m} \quad (8.19)$$

When $m$ increases, the $S$ and IL of the $GHZ_m$ state remain unchanged as in Eq. (8.15), while the S and IL of the $W_m$ state decrease and approach 0 (like a pure state) as shown in Fig. (8.1c). The $GHZ_m$ state ($m > 2$) has only one maximally entangled pair of qubits, its MEE is ln2, and its MEI is 1/2 as in Eq. (6.24), which is completely lost by measuring any qubit in CB. Meanwhile, the $W_m$ state has $m - 1$ maximally entangled pairs; measuring one qubit will only remove the entanglement info of just one such pair. Therefore, when $m$ increases, the $W_m$ state becomes increasingly more robust than the $GHZ_m$ state.

## 9. The Werner state and the Minimal Entanglement Entropy

So far, the initial state of all cases we have discussed is pure. Now, let us look at one simple (but very famous) mixed state, the 2-qubit Werner State [10] (also see [32, Sec.3]). Its density matrix can be written as:

$$\rho_\alpha = \alpha \, | \Psi^- \rangle \langle \Psi^- | + (1 - \alpha) \, I_2 \otimes I_2/4, \quad -1/3 \leq \alpha \leq 1. \quad (9.1)$$

Here $| \Psi^- \rangle$ is a Bell state in Eq. (5.3), α is the Werner parameter, and $I_2$ is a 2×2 identity matrix. The density matrix can be diagonalized, and its eigenvalues are [32, Sec.3]:

$$\lambda_1 = \lambda_2 = \lambda_3 = (1 - \alpha)/4, \; \lambda_4 = (1 + 3\alpha)/4. \quad (9.2)$$

From Eq. (3.4), the QE is given by:

$$S_\alpha \triangleq S(\rho_\alpha) = -3\left(\frac{1-\alpha}{4}\right) \ln \frac{1-\alpha}{4} - \frac{1+3\alpha}{4} \ln \frac{1+3\alpha}{4} \quad (9.3)$$

Assume that the initial state is in Bell state $|\Psi^-\rangle$ with $S_b = 0$ and the Werner state is the final state after certain irreversible operations, then IR and IL are given by Eq. (3.15) as

$$iR(S_\alpha) = iR(S_\alpha, 0) = \prod_{k=0}^{3} \eta_k^{\eta_k} = \left(\frac{1-\alpha}{4}\right)^{3(1-\alpha)/4} \left(\frac{1+3\alpha}{4}\right)^{(1+3\alpha)/4}, \; iL(S_\alpha) = 1 - iR(S_\alpha) \quad (9.4)$$

The graphs of Eq. (9.3-4) as functions of α are shown in Fig. (9.1.**a**). If α = 0, the state is maximally mixed; it has maximal entropy (ln 4 ~1.38) and a minimal retrievability of 25%. At α = 1, the state becomes a pure Bell state with zero entropy and 100% retrievability.

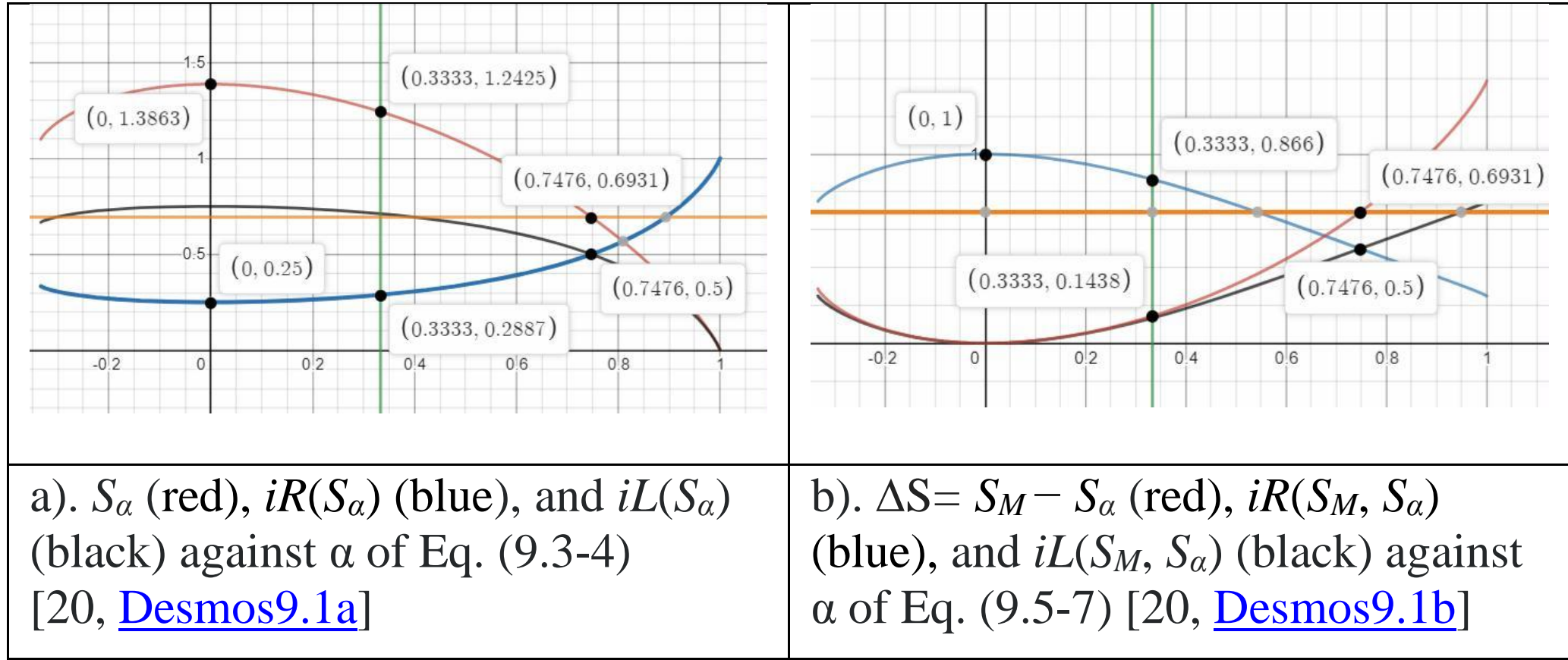

| a). $S_\alpha$ (red), $iR(S_\alpha)$ (blue), and $iL(S_\alpha)$ (black) against α of Eq. (9.3-4) [20, Desmos9.1a] | b). $\Delta S = S_M - S_\alpha$ (red), $iR(S_M, S_\alpha)$ (blue), and $iL(S_M, S_\alpha)$ (black) against α of Eq. (9.5-7) [20, Desmos9.1b] |
|---|---|

Fig. 9.1: a) From Bell to Werner; b) From Werner to MMS

Now suppose Werner state is the initial state, and the final state is a MMS, with maximal entropy $S_M = \ln 4$, as in Eq. (7.5). The corresponding IR and IL for $S_M$ are:

$$S_M = \ln 4 = 1.386, \qquad iR(S_M) = e^{-\ln 4} = 1/4 = 0.25, \qquad iL(S_M) = 3/4 = 0.75 \tag{9.5}$$

Using Eq. (3.13), we obtain the comparative retrievability as shown in Fig (9.1.b):

$$iR(S_M, S_\alpha) = e^{S_\alpha - S_M} = \frac{iR(S_M)}{iR(S_\alpha)} = \frac{1}{4}\left(\frac{1-\alpha}{4}\right)^{-3(1-\alpha)/4} \cdot \left(\frac{1+3\alpha}{4}\right)^{-(1+3\alpha)/4} \tag{9.6}$$

$$iL(S_M, S_\alpha) = 1 - iR(S_M, S_\alpha) \tag{9.7}$$

At α = 0, both the initial and final states are maximally mixed, and there is no change in entropy and no loss of information; at α = 1, the initial state is a pure state, the final state is an MMS, and the loss of information is maximal: $iL = 1 - iR = 1 - 1/4 = 3/4$.

Based on PPT (Positive Partial Transport) and Reduction Criterion, a Werner state $\rho_\alpha$ is separable if and only if $\alpha \le 1/3$. In contrast, based on von Neumann entropy inequality (vNEI), the necessary condition for its separability is $S_\alpha \ge \ln 2 \triangleq S(\rho_{vNEI})$ or $\alpha \le 0.7476$ [30, Sec.3].

The entropy and retrievable info for the four characteristic α values are shown in Table 9.1. To understand the data in the last row, please recall that, by definition, we have:

$$S(\rho_b \to \rho_M) = S(\rho_b \to \rho_\alpha) + S(\rho_\alpha \to \rho_M), \; iR(\rho_b \to \rho_M) = iR(\rho_b \to \rho_\alpha) \cdot iR(\rho_\alpha \to \rho_M) \tag{9.8}$$

| α: | 0 (MMS) | | 1/3 (PPT) | | 0.7476 (vNEI) | | 1(Bell) | |
|---|---|---|---|---|---|---|---|---|
| | $\Delta S$ | $IR$ | $\Delta S$ | $IR$ | $\Delta S$ | $IR$ | $\Delta S$ | $IR$ |
| $\rho_b \to \rho_\alpha$ | ln4 = 1.386 | 1/4 | 1.242 | 0.2887 | ln2 = 0.6931 | 1/2 | 0 | 1 |
| $\rho_\alpha \to \rho_M$ | 0 | 1 | 0.144 | 0.866 | ln2 | 1/2 | ln 4 | 1/4 |
| $\rho_b \to \rho_M$ | ln4 | 1/4 | 1.386 | 0.25 | ln4 = 1.386 | 1/4 | ln 4 | 1/4 |

Table 9.1: Werner State: Entropy and IR at characteristic values of α

It is interesting to note that the Minimal Entanglement Entropy (MEE), defined in Eq. (6.25) for Bell tests, the *MEE* for GHZ state, given in Eq. (8.20), and the vNEI margin of separability for the Werner state in Fig. (9.1a) are all equal. So are the MEI:

$$S\,(\rho_{vNEI}) \triangleq MEE\,(\text{Werner}) = MEE\,(\text{Bell}) = MEE\,(\text{GHZ}_m) = \ln 2 = 0.6931 \qquad (9.12)$$

$$\therefore MEI\,(\rho_{vNEI}) \triangleq MEI\,(\text{Werner}) = MEI\,(\text{Bell}) = MEI\,(\text{GHZ}_m) = 1 - e^{-\ln 2} = 1/2 \qquad (9.13)$$

Hence, when the entangled pair of Bell, GHZ, or vNEI-marginal Werner states are measured, they have the same minimal entanglement entropy gain (ln2, irremovable) associated with the equal minimal entanglement info loss (50%), which is not recoverable.

## 10. Summary and Discussion

We investigated entropy gain and information loss by measuring various states, whether pure or mixed. We defined Retrievability (IR) and Information Loss (IL) as functions of the density matrix through the von Neumann Entropy. When a pure, unbiased 10-qubit state collapses into a maximally mixed state, it experiences the maximal loss of information (99.9%) and the highest gain in entropy, equivalent to the 10-bit classical Shannon entropy. We verified that entropy gain is greater than the minimal bound of the entropy uncertainty in the Bell test and teleportation. After measuring, entangled pairs of Bell, GHZ, or vNEI-marginal Werner states yield the same irremovable minimal entanglement entropy gain of ln2 and an equal unrecoverable minimal entanglement information loss of 50%. We noted, "Gain in entropy always means loss of information, and nothing more” [33, Sec. 4.2].

However, sharing recorded knowledge between the observers in the Bell test [4, p111] [5] (as well as in the delayed choice scenario [6-7]) could recover certain quantum information even years later. In that case, the associated quantum entropy must be removable due to “the intervention of intelligent beings” [24]. This may suggest an explanation related to the Bell nonlocality [34]: No data exchange means additional information loss, additional entropy gain, and neither a violation of Bell’s inequality nor a spooky action at a distance. Having data exchanged later leads to recovering the missing information, removing the extra entropy gain, and violating Bell’s inequality. However, data exchange cannot go faster than light, so there is no spooky action at a distance either!

We adhere to the Copenhagen interpretation of measurements [15, 35]. Despite facing criticism, it provides a relatively simple framework for linking the irreversible nature of measurement with the probabilistic predictions of the Born rule. Engaging topics, such as whether the lost information transfers to alternate worlds (Many-Worlds interpretation, [36]) or the environment as a whole (No-Hiding theorem, [37]), fall outside the scope of this article.

**Abbreviations**:

| | |
|---|---|
| CB | Computational Basis |
| EOM | Electro-Optic Modulator |
| EE | Entanglement Entropy |
| EI | Entanglement Information |
| GHZ | Greenberger-Horne-Zeilinger |
| IL | Information Loss |
| IR | Information Retrievability |
| MMS | Maximally Mixed State |
| MEE | Minimal Entanglement Entropy |
| MEI | Minimal Entanglement Information |
| MQI | Mutual Quantum Information |
| PBN | Probability Bracket Notation |
| PMF | Probability Mass Function |
| QE | Quantum Entropy |
| QLC | Quantum Logic Circuit |
| VNE | von Neumann Entropy |